\documentstyle[twoside,psfig]{article}

\catcode`\@=11
\long\def\@makefntext#1{
\protect\noindent \hbox to 3.2pt {\hskip-.9pt  
$^{{\eightrm\@thefnmark}}$\hfil}#1\hfill}		%CAN BE USED 

\def\@makefnmark{\hbox to 0pt{$^{\@thefnmark}$\hss}}	%ORIGINAL 
	
\def\ps@myheadings{\let\@mkboth\@gobbletwo
\def\@oddhead{\hbox{}
\rightmark\hfil\eightrm\thepage}   
\def\@oddfoot{}\def\@evenhead{\eightrm\thepage\hfil
\leftmark\hbox{}}\def\@evenfoot{}
\def\sectionmark##1{}\def\subsectionmark##1{}}

\oddsidemargin=\evensidemargin
\newcounter{sectionc}\newcounter{subsectionc}\newcounter{subsubsectionc}
\renewcommand{\section}[1] {\vspace{12pt}\addtocounter{sectionc}{1} 
\setcounter{subsectionc}{0}\setcounter{subsubsectionc}{0}\noindent 
	{\tenbf\thesectionc. #1}\par\vspace{5pt}}
\renewcommand{\subsection}[1] {\vspace{12pt}\addtocounter{subsectionc}{1} 
	\setcounter{subsubsectionc}{0}\noindent 
	{\bf\thesectionc.\thesubsectionc. {\kern1pt \bfit #1}}\par\vspace{5pt}}
\renewcommand{\subsubsection}[1] {\vspace{12pt}\addtocounter{subsubsectionc}{1}
	\noindent{\tenrm\thesectionc.\thesubsectionc.\thesubsubsectionc.
	{\kern1pt \tenit #1}}\par\vspace{5pt}}

\newcounter{appendixc}
\newcounter{subappendixc}[appendixc]
\newcounter{subsubappendixc}[subappendixc]
\renewcommand{\thesubappendixc}{\Alph{appendixc}.\arabic{subappendixc}}
\renewcommand{\thesubsubappendixc}
	{\Alph{appendixc}.\arabic{subappendixc}.\arabic{subsubappendixc}}

\renewcommand{\appendix}[1] {\vspace{12pt}
        \refstepcounter{appendixc}
        \setcounter{figure}{0}
        \setcounter{table}{0}
        \setcounter{lemma}{0}
        \setcounter{theorem}{0}
        \setcounter{corollary}{0}
        \setcounter{definition}{0}
        \setcounter{equation}{0}
        \renewcommand{\thefigure}{\Alph{appendixc}.\arabic{figure}}
        \renewcommand{\thetable}{\Alph{appendixc}.\arabic{table}}
        \renewcommand{\theappendixc}{\Alph{appendixc}}
        \renewcommand{\thelemma}{\Alph{appendixc}.\arabic{lemma}}
        \renewcommand{\thetheorem}{\Alph{appendixc}.\arabic{theorem}}
        \renewcommand{\thedefinition}{\Alph{appendixc}.\arabic{definition}}
        \renewcommand{\thecorollary}{\Alph{appendixc}.\arabic{corollary}}
        \renewcommand{\theequation}{\Alph{appendixc}.\arabic{equation}}
        \noindent{\tenbf Appendix \theappendixc #1}\par\vspace{5pt}}
\newcommand{\subappendix}[1] {\vspace{12pt}
        \refstepcounter{subappendixc}
        \noindent{\bf Appendix \thesubappendixc. {\kern1pt \bfit #1}}
	\par\vspace{5pt}}
\newcommand{\subsubappendix}[1] {\vspace{12pt}
        \refstepcounter{subsubappendixc}
        \noindent{\rm Appendix \thesubsubappendixc. {\kern1pt \tenit #1}}
	\par\vspace{5pt}}

\newcommand{\textlineskip}{\baselineskip=13pt}
\newcommand{\smalllineskip}{\baselineskip=10pt}

\def\eightcirc{
\begin{picture}(0,0)
\put(4.4,1.8){\circle{6.5}}
\end{picture}}
\def\eightcopyright{\eightcirc\kern2.7pt\hbox{\eightrm c}} 

\newcommand{\publisher}[2]{{\begin{center}\footnotesize\smalllineskip 
	Received #1\\
	Revised #2
	\end{center}
	}}

\def\abstracts#1#2#3{{
	\centering{\begin{minipage}{4.5in}\footnotesize\baselineskip=10pt
	\parindent=0pt #1\par 
	\parindent=15pt #2\par
	\parindent=15pt #3
	\end{minipage}}\par}} 

\newcommand{\bibit}{\nineit}

\renewenvironment{thebibliography}[1]
	{\frenchspacing
	 \ninerm\baselineskip=11pt
	 \begin{list}{\arabic{enumi}.}
	{\usecounter{enumi}\setlength{\parsep}{0pt}
	 \setlength{\leftmargin 12.7pt}{\rightmargin 0pt} %FOR 1--9 ITEMS
	 \setlength{\itemsep}{0pt} \settowidth
	{\labelwidth}{#1.}\sloppy}}{\end{list}}

\newcounter{itemlistc}
\newcounter{romanlistc}
\newcounter{alphlistc}
\newcounter{arabiclistc}

\newcommand{\fcaption}[1]{
        \refstepcounter{figure}
        \setbox\@tempboxa = \hbox{\footnotesize Fig.~\thefigure. #1}
        \ifdim \wd\@tempboxa > 5in
           {\begin{center}
        \parbox{5in}{\footnotesize\smalllineskip Fig.~\thefigure. #1}
            \end{center}}
        \else
             {\begin{center}
             {\footnotesize Fig.~\thefigure. #1}
              \end{center}}
        \fi}

\newcommand{\tcaption}[1]{
        \refstepcounter{table}
        \setbox\@tempboxa = \hbox{\footnotesize Table~\thetable. #1}
        \ifdim \wd\@tempboxa > 5in
           {\begin{center}
        \parbox{5in}{\footnotesize\smalllineskip Table~\thetable. #1}
            \end{center}}
        \else
             {\begin{center}
             {\footnotesize Table~\thetable. #1}
              \end{center}}
        \fi}

\def\@citex[#1]#2{\if@filesw\immediate\write\@auxout
	{\string\citation{#2}}\fi
\def\@citea{}\@cite{\@for\@citeb:=#2\do
	{\@citea\def\@citea{,}\@ifundefined
	{b@\@citeb}{{\bf ?}\@warning
	{Citation `\@citeb' on page \thepage \space undefined}}
	{\csname b@\@citeb\endcsname}}}{#1}}

\newif\if@cghi
\def\cite{\@cghitrue\@ifnextchar [{\@tempswatrue
	\@citex}{\@tempswafalse\@citex[]}}
\def\citelow{\@cghifalse\@ifnextchar [{\@tempswatrue
	\@citex}{\@tempswafalse\@citex[]}}
\def\@cite#1#2{{$\null^{#1}$\if@tempswa\typeout
	{IJCGA warning: optional citation argument 
	ignored: `#2'} \fi}}

\def\pmb#1{\setbox0=\hbox{#1}
	\kern-.025em\copy0\kern-\wd0
	\kern.05em\copy0\kern-\wd0
	\kern-.025em\raise.0433em\box0}

\def\fnt#1#2{\footnotetext{\kern-.3em
	{$^{\mbox{\scriptsize #1}}$}{#2}}}

\def\@makefnmark{\hbox to 0pt{$^{\@thefnmark}$\hss}}	%ORIGINAL 
	
\def\ps@myheadings{%
    \let\@oddfoot\@empty\let\@evenfoot\@empty
    \def\@evenhead{\slshape\leftmark\hfil}%       %EVEN PAGE
    \def\@oddhead{\hfil{\slshape\rightmark}}%     %ODD PAGE
    \let\@mkboth\@gobbletwo
    \let\sectionmark\@gobble
    \let\subsectionmark\@gobble
    }
\font\tenrm=cmr10
\font\tenit=cmti10 
\font\tenbf=cmbx10
\font\bfit=cmbxti10 at 10pt
\font\ninerm=cmr9
\font\nineit=cmti9

\font\eightrm=cmr8

\newtheorem{theorem}{\indent Theorem}
\newtheorem{lemma}{Lemma}
\newtheorem{corollary}{Corollary}
\def\qed{\hbox{${\vcenter{\vbox{			%HOLLOW SQUARE
   \hrule height 0.4pt\hbox{\vrule width 0.4pt height 6pt
   \kern5pt\vrule width 0.4pt}\hrule height 0.4pt}}}$}}

\begin{document}
\setlength{\textheight}{7.7truein}  %for 2nd page onwards

\thispagestyle{empty}

\markboth{\protect{\footnotesize\it Instructions for Typesetting
Manuscripts}}{\protect{\footnotesize\it Instructions for
Typesetting Manuscripts}}

\normalsize\textlineskip

\setcounter{page}{1}

%\copyrightheading{}		%{Vol.~0, No.~0 (2000) 000--000}

\vspace*{0.88truein}

%\fpage{1}
\centerline{\bf THERMODYNAMIC PROPERTIES OF THE 2N--PIECE}
\vspace*{0.035truein}
\centerline{\bf RELATIVISTIC  STRING}
%\footnote{For
%the title, try not to use more than 3 lines. Typeset the title
%in 10 pt Times Roman, uppercase and boldface.}

\vspace*{0.37truein}
\centerline{\footnotesize IVER BREVIK}
%\footnote{Typeset names in
%10 pt Times Roman, uppercase. Use the footnote to indicate the
%present or permanent address of the author.}}
\baselineskip=12pt
\centerline{\footnotesize\it Division of Applied Mechanics, Norwegian 
University of Science and Technology,} 
\centerline{\footnotesize\it  N-7491, Trondheim, Norway.}
\centerline{\footnotesize\it E-mail: iver.h.brevik@mtf.ntnu.no}
\vspace*{10pt}

\centerline{\footnotesize ANDREI A. BYTSENKO}
%\footnote{Typeset names in
%10 pt Times Roman, uppercase. Use the footnote to indicate the
%present or permanent address of the author.}}
\baselineskip=12pt
\centerline{\footnotesize\it Departamento de Fisica, Universidade
Estadual de Londrina,}
\centerline{\footnotesize\it Caixa Postal 6001, Londrina--Parana, Brazil}
\centerline{\footnotesize\it E-mail: abyts@uel.br}
%Country\footnote{State completely without abbreviations, the
%affiliation and mailing address, including country. Typeset in 8
%pt Times Italic.}}
\vspace*{10pt}

\centerline{\footnotesize ROGER SOLLIE}
%\footnote{Typeset names in
%10 pt Times Roman, uppercase. Use the footnote to indicate the
%present or permanent address of the author.}}
\baselineskip=12pt
\centerline{\footnotesize\it STATOIL Research Centre, N-7005 Trondheim, Norway} 
\centerline{\footnotesize\it E-mail: rsol@statoil.com}
%\baselineskip=10pt

%\vspace*{10pt}
%\centerline{\footnotesize SECOND AUTHOR}
%\baselineskip=12pt
%\centerline{\footnotesize\it Group, Laboratory, Address}
%\baselineskip=10pt
%\centerline{\footnotesize\it City, State ZIP/Zone, Country}
\vspace*{0.225truein}
\publisher{(received date)}{(revised date)}

\vspace*{0.21truein}

\abstracts{The thermodynamic free energy $F(\beta)$ is calculated for
a gas consisting of the transverse oscillations of a piecewise uniform
bosonic string. The string consists of $2N$ parts of equal length, of
alternating type I and type II material, and is relativistic in the
sense that the velocity of sound everywhere equals the velocity of
light. The present paper is a continuation of two earlier papers, one
dealing with the Casimir energy of a $2N$--piece string [I. Brevik and
R. Sollie (1997)], and another dealing with the thermodynamic
properties of a string divided into two (unequal) parts [I. Brevik,
A. A. Bytsenko and H. B. Nielsen (1998)]. Making use of the Meinardus
theorem we calculate the asymptotics of the level state density, and
show that the critical temperatures in the individual parts are equal,
for arbitrary spacetime dimension $D$. If $D=26$, we find $\beta=
(2/N)\sqrt{2\pi /T_{II}}$, $T_{II}$ being the tension in part
II. Thermodynamic interactions of parts related to high genus ${\rm
g}$ is also considered.}{}{}

%\textlineskip			%) USE THIS MEASUREMENT WHEN THERE IS
%\vspace*{12pt}			%) NO SECTION HEADING

\vspace*{1pt}\textlineskip	%) USE THIS MEASUREMENT WHEN THERE IS

\section{Introduction}

Whereas the bosonic string of length $L$ in $D$-dimensional spacetime
is assumed to be uniform, the {\it composite} string is imagined to
consist of two or more uniform pieces. In a Casimir context, such a
model was introduced in 1990 \cite{brevniels90}. The string was
assumed to divided into two pieces, of length $L_I$ and $L_{II}$, and
the model was relativistic in the sense that the velocity of sound was
everywhere required to be equal to the velocity of light. With this
contraint imposed on the model, the Casimir energy of the string,
i.e., the zero--point energy associated with its discontinuity
properties, was easily calculable as a function of the length ratio
$s=L_{II}/L_I$. Later, various aspects of the relativistic piecewise
uniform string model were studied
\cite{bayin96,berntsen97,brevik94,brevik95,brevik96,brevik97,brevik98,BBG,brevik01,li91,lu}.
One may note, for instance, the paper of Lu and Huang \cite{lu} in which
the model finds application in relation to the Green-Schwarz
superstring.

The present paper focuses attention on the thermodynamic free energy
$F(\beta)$ at inverse temperature $\beta=1/T$ of a $2N-$ piece string,
made up of $2N$ parts of equal length, of alternating type I and type
II material. The model is relativistic, in the sense explained
above. In an earlier paper \cite{brevik97} we developed the Casimir
theory for a string of this type, whereas in another paper
\cite{brevik98} we considered the free energy for the case where the
string consists of {\it two} pieces only, i.e., the model of
Ref.~\cite{brevniels90}. The calculation of $F(\beta)$ for a
$2N-$ piece string has to our knowledge not been undertaken before.  It
turns out, similarly as in Ref.~\cite{brevik01}, that the Meinardus
theorem \cite{meinardus59,meinardus61,andrews} is powerful, allowing
us to find the asymptotics of the level state density. Using this we
find, for a general spacetime dimension $D$, that the critical
(Hagedorn) temperatures for the two kinds of pieces are the same. When
$D=26$, the common spacetime dimension for a bosonic string, we find
$\beta_c=(2/N)\sqrt{2\pi /T_{II}},~T_{II}$ being the tension in region
II.  This result is derived in Section 6. In Section 7, we comment
upon the thermodynamic properties of the composite string for
arbitrary genus {\rm g}.

\section{Resum\'{e} of the 2N--Piece String Theory}

Assume, as mentioned, that the string of total length $L$ is divided 
into $2N$ equally large pieces, of alternating type I and type II 
material; see Fig.~1. The string is relativistic, in the sense that the 
velocity of sound is everywhere equal to the velocity of light,
$v_s=\sqrt{T_I/\rho_I}=\sqrt{T_{II}/\rho_{II}}=c$,
where $T_I, T_{II}$ are the tensions and $\rho_I, \rho_{II}$ the mass 
densities in the two pieces. We will study the transverse oscillations 
$\psi=\psi(\sigma,\tau)$ of the string; $\sigma$ denoting as usual the 
position coordinate and $\tau$ the time coordinate of the string. We can 
thus write in the two regions
$$
\psi_I=\xi_I e^{i\omega (\sigma-\tau)}+\eta_I e^{-i\omega (\sigma+\tau)}, 
$$

\begin{equation}
\psi_{II}=\xi_{II} e^{i\omega (\sigma-\tau)}+\eta_{II} e^{-i\omega 
(\sigma+\tau)}
\mbox{,}
\label{1}
\end{equation}
\\
where $\xi$ and $\eta$ are constants. The junction conditions are that 
$\psi$ itself as well as the transverse elastic force 
$T\partial \psi /\partial \sigma$ are continuous, i.e.,

\begin{equation}
\psi_I=\psi_{II},\,\,\,\,\,\,\,\,\,\,
T_I\partial \psi_I/\partial \sigma =T_{II}
\partial \psi_{II}/\partial \sigma
\mbox{,} 
\label{2}
\end{equation}
at each of the $2N$ junctions. We define $x$ as the tension ratio,
$x=T_I/T_{II}$, and define also the symbols $p_N$ and $\alpha$ by
$p_N=\omega L/N$,\,\,$\alpha= (1-x)/(1+x)$. Now introduce the matrix 
$\Lambda$,

\begin{equation}
\Lambda(\alpha,p_N)=\left( \begin{array}{cc}
a  & b \\
b^*& a^*
\end{array}
\right),
\end{equation}
\label{3}
with
\begin{equation}
a=e^{-ip_N}-\alpha^2,~~~~b=\alpha(e^{-ip_N}-1).
\end{equation}
\label{4}
%%%%%%%%%%%%%%%%%%%%%%%%%%%%%%%%%%%%%%%%%%%%%%%
% Figure 1
\begin{figure}
\centerline{
\psfig{file=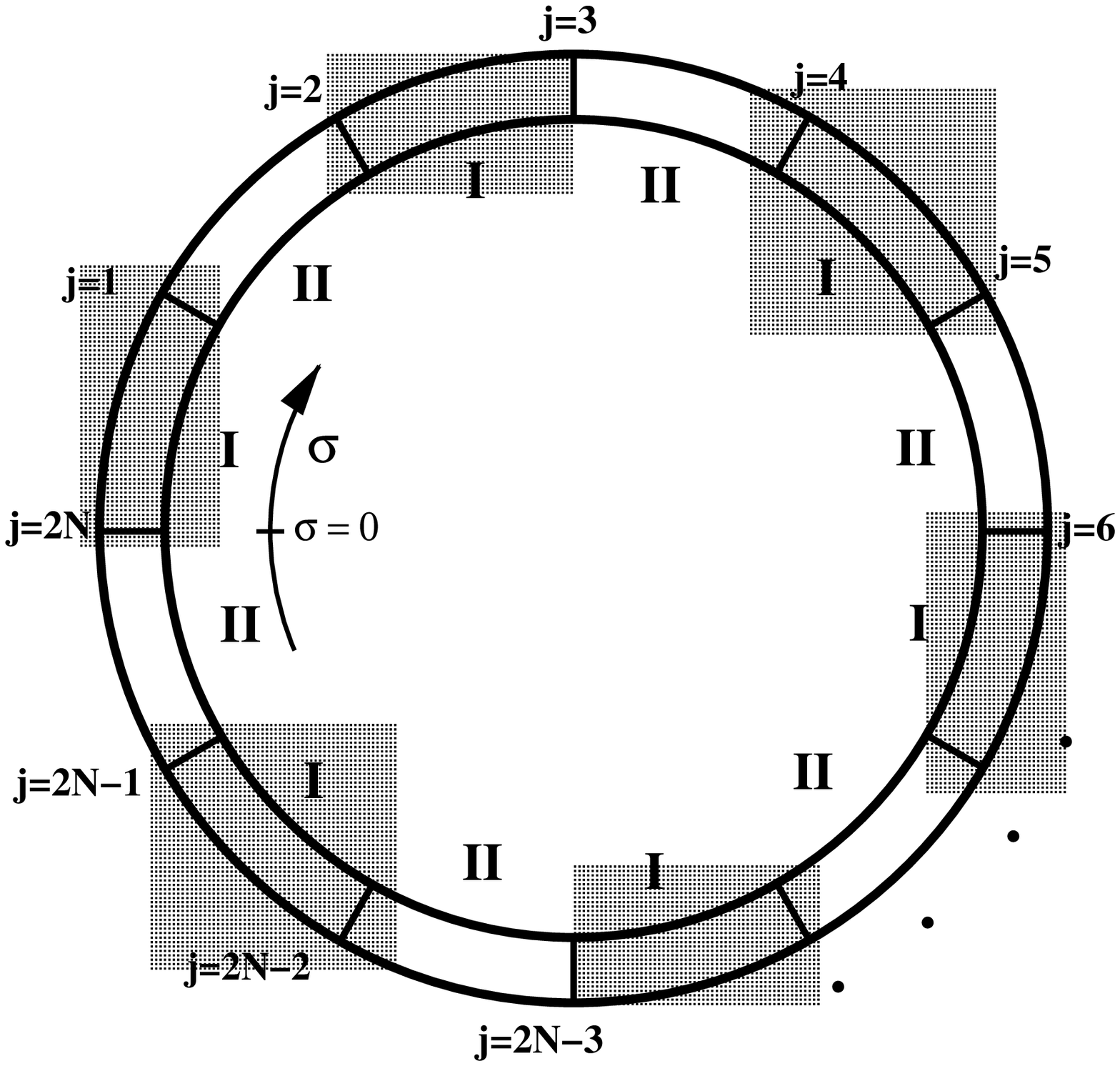,height=7.0cm} }
\fcaption{Sketch of the composite $2N$ string, when $N=6$.}
\end{figure}
%%%%%%%%%%%%%%%%%%%%%%%%%%%%%%%%%%%%%%%%%%%%%%%

Then, as shown in \cite{brevik97}, the eigenfrequencies $\omega$ are 
determined from the equation
\begin{equation}
{\it Det} \left[ (1-\alpha^2)^{-N}\Lambda^N(\alpha,p_N)-1 \right]=0.
\end{equation}
\label{5}
For practical purposes it is convenient to reformulate the condition (5). 
Let us define two new quantities $\lambda_\pm$:

\begin{equation}
\lambda_\pm (p_N)=\cos p_N-\alpha^2 \pm \left[ (\cos p_N-\alpha^2)^2-
(1-\alpha^2)^2\right]^{1/2}.
\end{equation}
\label{6}
Then, we can re--express the condition (5) as \cite{brevik97}
\begin{equation}
\lambda_+^N+\lambda_-^N=2(1-\alpha^2)^N.
\end{equation}
\label{7}
We can now make use of the following recursion formula for the quantity 
$S_N \equiv \lambda_+^N+\lambda_-^N$:

\begin{equation}
S_N=2(\cos p-\alpha^2)S_{N-1}-(1-\alpha^2)^2 S_{N-2}, ~~N \geq 2,
\end{equation}
\label{8}
in which it is assumed that $\omega L/N$ is constant, at all recursive 
steps. The initial values of $S_N$ are 
$S_0=2,~S_1=\lambda_++\lambda_-=2(\cos p-\alpha^2).$

Assume now that $L=\pi$, in conformity with usual practice. 
Thus $p_N=\pi \omega/N$. We let $X^\mu(\sigma,\tau)$, with 
$\mu=0,1,2,.., (D-1)$, specify the coordinates on the world sheet. For each 
of the eigenvalue branches determined by the dispersion equation (5) we 
can write $X^\mu$ on the form

\begin{equation}
X^\mu=x^\mu+\frac{p^\mu \tau}{\pi {T_0}}+X_I^\mu,~~{\rm region ~I},
\end{equation}
\label{9}
\begin{equation}
X^\mu=x^\mu+\frac{p^\mu \tau}{\pi {T_0}}+X_{II}^\mu,~~{\rm region ~II},
\end{equation}
\label{10}
where $x^\mu$ is the centre--of--mass position, $p^\mu$ the total momentum 
of the string, and ${T_0}=\frac{1}{2}(T_I+T_{II})$
is the mean tension. Further, $X_I^\mu$ and $X_{II}^\mu$ are decomposed 
into oscillator coordinates,

\begin{equation}
X_I^\mu=\frac{i}{2}{\ell}_s \sum_{n \neq 0} \frac{1}{n}
\left[ \alpha_{nI}e^{i\omega (\sigma-\tau)}+\tilde{\alpha}_{nI}
e^{-i\omega(\sigma+\tau)} \right],
\end{equation}
\label{11}

\begin{equation}
X_{II}^\mu=\frac{i}{2}{\ell}_s\sum_{n\neq 0}\frac{1}{n}\left[ 
\alpha_{nII}e^{i\omega (\sigma-\tau)}+\tilde{\alpha}_{nII}e^{-i\omega 
(\sigma+\tau)} \right].
\end{equation}
\label{12}
Here,\, ${\ell}_s$ is the fundamental string length, unspecified so far, and 
$\alpha_n, \tilde{\alpha}_n$ are oscillator coordinates of the right--and 
left--moving waves, respectively. A characteristic property of the 
composite string is that the oscillator coordinates have to be specified 
for each of the various branches determined by Eq.~(5). This makes the 
handling of the formalism complicated, in general. A significant 
simplification can be obtained if, following Ref.~\cite{brevik98}, we 
limit ourselves to the case of extreme string ratios only. Since 
$\alpha$ occurs quadratically in Eqs.~(6) and (7), the eigenvalue 
spectrum has to be invariant inder the transformation 
$x \rightarrow 1/x$. It is sufficient, therefore, to consider the tension 
ratio interval $0<x \leq 1$ only. The case of extreme tensions corresponds to
$x \rightarrow  0$.
We consider only this case in the following.

\section{The Case of Extreme Tensions}

We assume that $T_{II}$  has a finite value, so that the limiting case 
$x \rightarrow 0$ corresponds to $T_I \rightarrow 0$. Thus 
${T}_0\rightarrow (1/2)T_{II}$. Since now $\alpha \rightarrow 1$ 
we get from Eq.~(6) $\lambda_-=0,~\lambda_+=\cos p_N-1$, so we obtain 
from Eq.~(7) the remarkable simplification that all the eigenfrequency 
branches degenerate into one single branch determined by $\cos p_N=1$. 
That is, the eigenvalue spectrum becomes

\begin{equation}
\omega_n=2Nn,~~~n=\pm 1, \pm 2, \pm 3,...
\label{13}
\end{equation}
Then, choosing the fundamental length equal to ${\ell}_s= (\pi T_I)^{-1/2}$,
we can write the expansion (11) in region I as (subscript I on the 
$\alpha_n$'s omitted)

\begin{equation}
X_I^\mu=\frac{i}{2\sqrt{\pi T_I}}\sum_{n\neq 0}\frac{1}{n}\left[ 
\alpha_n^\mu e^{2iNn(\sigma-\tau)}+\tilde{
\alpha}_n^\mu e^{-2iNn(\sigma+\tau)} \right].
\label{14}
\end{equation}
\\
The junction conditions (2) permit all waves to propagate from region 
I to region II. When $x\rightarrow 0$, they reduce to the equations

\begin{equation}
\xi_I+\eta_I=2\xi_{II}=2\eta_{II},
\label{15}
\end{equation}
which show that the right-- and left-- moving amplitudes $\xi_I$ and 
$\eta_I$ in region I can be chosen freely and that the amplitudes 
$\xi_{II}, \eta_{II}$ in region II are thereafter fixed. This means, 
in oscillator language, that $\alpha_n^\mu$ and $\tilde{\alpha}_n^\mu$ 
can be chosen freely. The expansion in region II can in view of Eq.~(15) 
be written as

\begin{equation}
X_{II}^\mu=\frac{i}{2\sqrt{\pi T_I}}\sum_{n\neq 0}\frac{1}{n}\gamma_n^\mu 
e^{-2iNn\tau}\cos(2Nn\sigma),
\label{16}
\end{equation}
where we have defined $\gamma_n^\mu$ as

\begin{equation}
\gamma_n^\mu=\alpha_n^\mu+\tilde{\alpha}_n^\mu,~~~n \neq 0.
\label{17}
\end{equation}
The oscillations in region II are thus standing waves. This is the same 
kind of behaviour as that found for the two-piece string \cite{brevik98}.

The action of the string is

\begin{equation}
S=-\frac{1}{2}\int d\tau d\sigma T(\sigma) \eta^{\alpha \beta}
\partial_\alpha X^\mu \partial_\beta X_\mu,
\label{18}
\end{equation}
where $\alpha, \beta =0,1$ and $T(\sigma)=T_I$ in region I,\, 
$T(\sigma)=T_{II}$ in region II. The momentum conjugate to $X^\mu$ is 
$P^\mu(\sigma)=T(\sigma)\dot{X}^\mu$, and the Hamiltonian is accordingly

\begin{equation}
H=\int_0^\pi \left[ P_\mu(\sigma)\dot{X}^\mu- {\cal L} \right] d\sigma=
\frac{1}{2}
\int_0^\pi T(\sigma)(\dot{X}^2+X'^2)d\sigma,
\label{19}
\end{equation}
where ${\cal L}$ is the Lagrangian.

As for the constraint equation for the string, some care has to be taken. 
Conventionally, in the classical theory for the uniform string the constraint 
equation reads $T_{\alpha \beta}=0$, $T_{\alpha \beta}$ being the 
energy--momentum tensor. As discussed in Ref.~\cite{brevik98} the situation 
is here more complicated, since the junctions restrict the freedom one has 
to take the variations $\delta X^\mu$. We thus have to replace the strong 
condition $T_{\alpha \beta}=0$ by a weaker condition, and the most natural 
choice, which we will adopt, is to to impose that $H=0$ when applied to the 
physical states.

Let us introduce lightcone coordinates, $\sigma^-=\tau-\sigma$ and 
$\sigma^+=\tau+\sigma$. The derivatives conjugate to $\sigma^\mp$ are 
$\partial_\mp=\frac{1}{2}(\partial_\tau \mp \partial_\sigma)$. In region I,

$$\partial_-X^\mu=\frac{N}{\sqrt{\pi T_I}}\sum_{-\infty} ^\infty 
\alpha_n^\mu e^{2iNn(\sigma-\tau)}, 
$$

\begin{equation}
\partial_+X^\mu=\frac{N}{\sqrt{\pi T_I}}\sum_{-\infty}^\infty 
\tilde{\alpha}_ne^{-2iNn(\sigma+\tau)},
\label{20}
\end{equation}
and in region II

\begin{equation}
\partial_\mp X^\mu=\frac{N}{2\sqrt{\pi T_I}}\sum_{-\infty}^\infty 
\gamma_n^\mu e^{\pm 2in(\sigma \mp\tau)},
\label{21}
\end{equation}
where we have defined

\begin{equation}
\alpha_0^\mu=\tilde{\alpha}_0^\mu=\frac{p^\mu}{NT_{II}}
\sqrt{\frac{T_I}{\pi}}, \,\,\,\,\,\,\,\,\,\,\,\,
\gamma_0^\mu=2\alpha_0^\mu.
\label{22}
\end{equation}
Inserting these expressions into the Hamiltonian

$$
H=\int_0^\pi T(\sigma)\left( \partial_-X\cdot \partial_-X+\partial_
+X\cdot \partial_+X \right) d\sigma
$$

$$
=NT_I\int_0^{\pi/(2N)}\left( \partial_-X \cdot \partial_- X+\partial_
+X \cdot \partial_+ X \right) d\sigma
$$

\begin{equation}
+NT_{II}\int_{\pi/(2N)}^{\pi/N}\left( \partial_-X\cdot \partial_-X
+\partial_+X\cdot \partial_+X \right) d\sigma 
\label{23}
\end{equation}

we get

\begin{equation}
H=\frac{1}{2}N^2\sum_{-\infty}^\infty (\alpha_{-n}\cdot \alpha_n 
+\tilde{\alpha}_{-n}\cdot \tilde{\alpha}_n) +
\frac{N^2}{4x}\sum_{-\infty}^\infty \gamma_{-n}\cdot \gamma_n .
\label{24}
\end{equation}
Now consider the expression for the square $M^2$ of the mass of the string. 
One must have $M^2=-p^\mu p_\mu$, as in the case of a uniform string 
\cite{green87}. We start from the constraint $H=0$ when applied to physical 
states, making use of Eq.~(24) in which we separate out the $n=0$ terms. 
Using that $\alpha_0\cdot \alpha_0=-M^2x/(\pi N^2 T_{II})$ according to 
Eq.~(22) we obtain in this way, when again observing that $x \ll 1$,

\begin{equation}
M^2=\pi N^2 T_{II}\sum_{n=1}^\infty \left[\alpha_{-n}\cdot \alpha_n
+\tilde{\alpha}_{-n}\cdot \tilde{\alpha}_n+\frac{1}{2x}\gamma_{-n}\cdot 
\gamma_n \right].
\label{25}
\end{equation}

\section{Quantization}

The momentum conjugate to $X^\mu$ is at any position on the string equal 
to $T(\sigma)\dot{X}^\mu$. We accordingly require the commutation rules 
in region I to be

\begin{equation}
T_I[\dot{X}^\mu (\sigma,\tau),X^\nu(\sigma',\tau)]=-i\delta (\sigma-\sigma')
\eta^{\mu\nu},
\label{26}
\end{equation}
and in region II

\begin{equation}
T_{II}[\dot{X}^\mu(\sigma,\tau), X^\nu(\sigma',\tau)]=-i\delta 
(\sigma-\sigma') \eta^{\mu\nu},
\label{27}
\end{equation}
$\eta^{\mu\nu}$ being the $D$-dimensional flat metric. The other commutators 
vanish. The quantities to be promoted to Fock state operators are 
$\alpha_{\mp n}$ and $\gamma_{\mp n}$. We insert the expansions for 
$X^\mu$ and $\dot{X}^\mu$ in regions I and II into Eqs.~(26) and (27) and 
make use of the effective relationship

\begin{equation}
\sum_{n=-\infty}^\infty e^{2iNn(\sigma -\sigma')}=2\sum_{n=-\infty}^\infty 
\cos 2Nn\sigma \cos 2Nn\sigma' 
\longrightarrow \frac{\pi}{N}\delta (\sigma -\sigma').
\label{28}
\end{equation}
We then get in region I

\begin{equation}
[\alpha_n^\mu, \alpha_m^\nu]=n\delta_{n+m,0}\eta^{\mu\nu},
\label{29}
\end{equation}
with a similar relation for $\tilde{\alpha}_n$. In region II,

\begin{equation}
[\gamma_n^\mu, \gamma_m^\nu]=4nx\,\delta_{n+m,0} \,\eta^{\mu\nu}.
\label{30}
\end{equation}
We introduce annihilation and creation operators by

$$
\alpha_n^\mu=\sqrt{n}\, a_n^\mu,~~~\alpha_{-n}^\mu=\sqrt{n}\,
a_n^{\mu \dagger},
$$

\begin{equation}
\gamma_n^\mu=\sqrt{4nx}\,c_n^\mu,~~~\gamma_{-n}^\mu=\sqrt{4nx}\,
c_n^{\mu \dagger},
\label{31}
\end{equation}
and find for $n \geq 1$ the standard form

$$
[a_n^\mu, a_m^{\nu \dagger}]=\delta_{nm}\eta^{\mu\nu}, 
$$

\begin{equation}
[c_n^\mu,c_m^{\nu\dagger}]=\delta_{nm}\eta^{\mu\nu}.
\label{32}
\end{equation}
These expressions are formally the same as those found for a two--piece 
string \cite{brevik98}. From Eq.~(24) we get, when separating out the 
$n=0$ term,

\begin{equation}
H=-\frac{M^2}{\pi T_{II}}+\frac{1}{2}N\sum_{n=1}^\infty \omega_n 
\left( a_n^\dagger \cdot a_n+\tilde{a}_n^\dagger\
\cdot \tilde{a}_n+2\,c_n^\dagger\cdot c_n \right).
\label{33}
\end{equation}
Here $a_n^\dagger \cdot a_n \equiv a_n^{\mu \dagger}a_{n\mu}$, and 
$\omega_n=2Nn$ as before.
From the condition $H=0$ we now get

\begin{equation}
M^2=\frac{1}{2}\pi NT_{II}\sum_{i=1}^{24}\sum_{n=1}^\infty \omega_n
\left( a_{ni}^\dagger a_{ni}+\tilde{a}_{ni}^\dagger \tilde{a}_{ni}
+2\,c_{ni}^\dagger c_{ni}- {\cal C} \right),
\label{34}
\end{equation}
where we have put $D=26$ and summed over the transverse 24 oscillator 
operators. Further, we have introduced a constant ${\cal C}$ in order 
to account for ordering ambiguities.

\section{Quantum Thermodynamics}

The constraint for the closed string (fat circles at Fig. 2), expressing the 
invariance of the theory in the region I under shifts of the origin of the 
co--ordinate, has the form

\begin{equation}
\sum_{i=1}^{24}\sum_{n=1}^{\infty}\omega_n \left[
a_{ni}^{\dagger}a_{ni} - \tilde{a}_{ni}^{\dagger}\tilde{a}_{ni}
\right]=0
\mbox{.}
\label{35}
\end{equation}
The commutation relations for the above operators are given by Eq. (32).
The mass of state (obtained by acting on the Fock vacuum $|0>$ with creation
operators) can be written as follows: 
$({\rm mass})^2\sim a_{n1}^{\dagger}...a_{ni}^{\dagger}c_{n1}^{\dagger}...
c_{ni}^{\dagger}|0>$.

\medskip
\medskip
\medskip
%%%%%%%%%%%%%%%%%%%%%%%%%%%%%%%%%%%%%%%%%%%%%%%
% Figure 2
\begin{figure}
\centerline{
\psfig{file=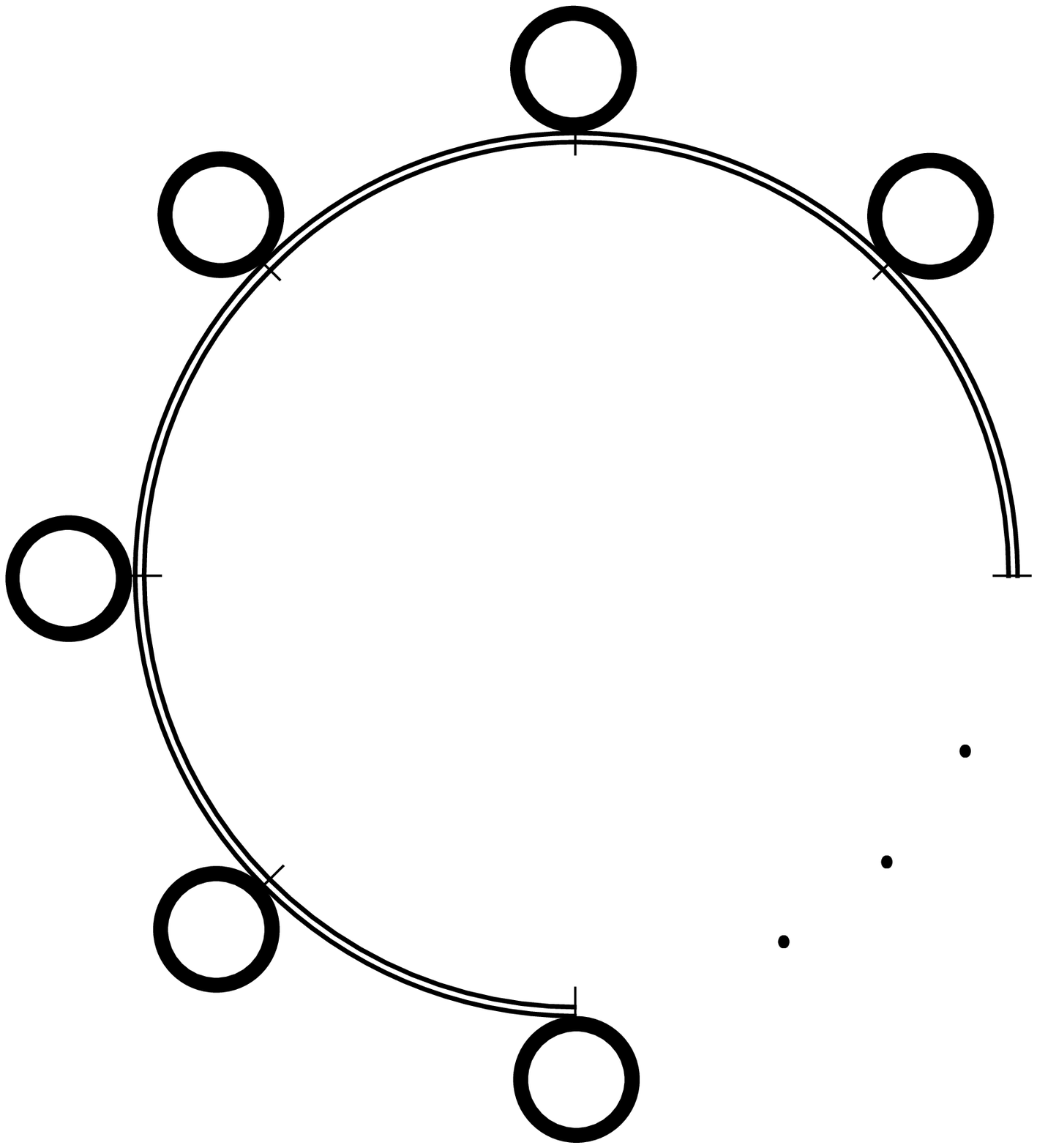,height=6.0cm}  }
\fcaption{Fat circles correspond to closed strings, double lines
correspond to open strings.}
\end{figure}
%%%%%%%%%%%%%%%%%%%%%%%%%%%%%%%%%%%%%%%%%%%%%%%
\medskip

Let us start with the discussion of the free energy in field theory
at non--zero temperature $T=\beta^{-1}$ (we put $k_B=1$). 
As usual the physical Hilbert space consists of all Fock
space states obeying the condition (35), which can be implemented by means
of the integral representation for Kronecker deltas. Thus the free energy
of the field content in the "proper time" representation becomes

$$
F(\beta) = {\cal F}(\beta=\infty)
- \pi(2\pi)^{-14}\int_0^{\infty}\frac{d\tau_2}{\tau_2^{14}}
\left[\theta_3\left(0|\frac{i\beta^2}{2\pi\tau_2}\right)-1\right]
{\rm Tr}\exp\left\{-\frac{\tau_2M^2}{2}\right\}
$$

\begin{equation}
\!\!\!\!\!\!\!\!\!\!\!\!\!\!\!\!\!\!\!\!\!\!\!\!\!\!\!
\times \int_{-\pi}^{\pi}\frac{d\tau_1}{2\pi}{\rm Tr}
\exp\left\{iN\tau_1\sum_{i=1}^{24}\sum_{n=1}^{\infty}\omega_n \left[
a_{ni}^{\dagger}a_{ni} - \tilde{a}_{ni}^{\dagger}\tilde{a}_{ni}
\right]\right\}
\mbox{,}
\label{36}
\end{equation}
where ${\cal F}(\beta=\infty)$ is the temperature independent part of 
$F(\beta)$ (the Casimir energy), while the second term in (36) presents
the temperature dependent part (the statistical sum).
Once the free energy has been found, the other thermodynamic quantities can
readily be calculated. For instance, the energy $U$ and the entropy $S$ of
the system are $U = (\partial/\partial \beta)(\beta F(\beta))\,,\,\,\,\,\,
S =  \beta^2 (\partial/\partial \beta) (F(\beta))$.

\section{The Critical Temperature}

First we consider some mathematical results on the asymptotics of the level 
degeneracy which leads to the asymptotics of the level state density.
Let  

\begin{equation}
{\cal G}(z)=\prod_{
n=1}^\infty[1-e^{-zn}]^{-a_n}=1+\sum_{n=1}^\infty \Xi (n)e^{-zn}
\mbox{,}
\label{37}
\end{equation}
be the generating function, where $\Re \,z >0$ and $a_n$ are
non--negative real numbers. Let us consider the associated Dirichlet
series 

\begin{equation}
{\cal D}(s)=\sum_{n=1}^\infty a_n n^{-s},\,\,\,\,\,\,\,\,
s=\sigma+it
\mbox{,}
\label{38}
\end{equation}
which converges for $0<\sigma<p$. We assume that ${\cal D}(s)$ can be 
analytically continued in the region $\sigma\geq-C_0 $  ($0<C_0<1$) and here 
${\cal D}(s)$
is analytic except for a pole of order one at $s=p$ with residue $A$.
Besides we assume that ${\cal D}(s)={\cal O}(|t|^{C_1})$ uniformly in
$\sigma\geq-C_0$ as $|t|\rightarrow\infty$, where $C_1$ is a fixed
positive real number. The following lemma 
\cite{meinardus59,meinardus61,andrews}
is useful with regard to the asymptotic properties of ${\cal G}(z)$ 
at $z=0$: 

\medskip
\par \noindent
\begin{lemma}
If ${\cal G}(z)$ and ${\cal D}(s)$ satisfy the above assumptions and 
$z=y+2\pi ix$ then

\begin{equation}
{\cal G}(z)=\exp{\{A\Gamma(p)\zeta_R(1+p) z^{-p}-{\cal D}(0){\rm log}
z+\frac{d}{ds}{\cal D}(s)|_{s=0}+{\cal O}(y^{C_0})\}}
\label{39}
\end{equation}
uniformly in $x$ as $y\rightarrow0$, provided $|\arg z|\leq\pi/2$ and 
$|x|\leq 1/2$. Moreover there exists a positive number $\varepsilon$ such that

\begin{equation}
{\cal G}(z)={\cal O}(\exp{\{A\Gamma(p)\zeta_R (1+p)
y^{-p}-Cy^{-\varepsilon}\}})
\mbox{,}
\label{40}
\end{equation}
uniformly in $x$ with $y^{\alpha}\leq|x|\leq1/2$ as $y\rightarrow0$, $C$ 
being a fixed real number and $\alpha=1+p/2-p\nu/4$, $0<\nu<2/3$.
\end{lemma}
\medskip
\medskip
The main result below follows from the Lemma and permits one to calculate 
the complete asymptotics of $\Xi(n)$. 

\medskip
\par \noindent
\begin{theorem}
{\rm (Meinardus \cite{meinardus59,meinardus61})}
For $n\rightarrow\infty$ one has 

\begin{equation}
\Xi (n)= C_pn^{k} \exp\left\{\frac{1+p}{p}
[A\Gamma(1+p)\zeta_R (1+p)]^{\frac1{1+p}}
n^{\frac{p}{1+p}}\right\}
\left(1+{\cal O}(n^{-k_1})\right)\,
\mbox{,}
\label{41}
\end{equation}

\begin{equation}
C_p=[A\Gamma(1+p)\zeta_R (1+p)]^{\frac{1-2{\cal D}(0)}{2(1+p)}}
\frac{\exp\left(\frac{d}{ds}{\cal D}(s)|_{s=0}\right)}
{[2\pi(1+p)]^{\frac12}} \,
\mbox{,}
\label{42}
\end{equation}

\begin{equation}
k=\frac{2{\cal D}(0)-p-2}{2(1+p)}\,\,,\,\,\,\,\,\,\,\,\,\,
k_1=\frac{p}{1+p}\min
\left(\frac{C_0}{p}-\frac{\nu}{4},\frac{1}{2}-\nu\right)\,
\mbox{.}
\label{43}
\end{equation}
\end{theorem}
\medskip
\medskip
Coming back to composite string problem note that the generation function
has the form

$$
W(\beta)= {\rm Tr}\left[e^{-\beta M^ 2}\right]_{({\cal C}=0)} =
W^{(I)}(\beta)W^{(II)}(\beta)
$$

\begin{equation}
\equiv 
\prod_{n=1}^{\infty}\left[1-e^{-n\beta Q(N)}\right]^{-48}
\prod_{n=1}^{\infty}\left[1-e^{-2n\beta Q(N)}\right]^{-24}
\label{44}
\mbox{,}
\end{equation}
where $Q(N)= \pi T_{II}N^2$.
Some remarks are in order. Taking into account $a_n=D-2$ (or $a_n=2(D-2)$
in the case of region II), 
we have $p=1$. In fact Eq. (38) gives the Riemann zeta function.
Therefore, from the Meinardus theorem, Eqs. (41)--(43), it follows that

\begin{equation}
\Xi^{(I)} (n)= C_1^{(I)}n^{k^{(I)}} 
\exp\left\{\pi\sqrt{\frac{4n\pi (D-2)}{3Q(N)}}\right\}
(1+{\mathcal O}(n^{-k_1}))\,
\mbox{,}
\label{45}
\end{equation}

\begin{equation}
\Xi^{(II)} (n)= C_1^{(II)}n^{k^{(II)}} 
\exp\left\{\pi \sqrt{\frac{2n(D-2)}{3Q(N)}}\right\}
(1+{\mathcal O}(n^{-k_1}))\,
\mbox{.}
\label{46}
\end{equation}
Using the mass formula $M^2=n$ (for the sake of simplicity we assume a
tension parameter, with dimensions of $(\rm{mass})^{p+1}$, equal to $1$) we
find for the number of bosonic string states of mass $M$ to $M+dM$

\begin{equation}
\nu(M)dM \simeq 2C_1M^{\frac{1-D}{2}}\exp(bM)dM
\,,\,\,\,\,\,\,\,\, b=\pi\sqrt{\frac{D-2}{3Q(N)}}\,\,
\mbox{.}
\label{47}
\end{equation}
One can show that the constant $b$ is the inverse of the Hagedorn
temperature. 

For the closed bosonic string in the region I the 
constraint $N_0 =\tilde{N}_0$ should be taking into account, 
where $N_0$ is a 
number operator related to $M^2$. As a result in Eq. (45) the total 
degeneracy of the level $n$ is simply the square of $\Xi^{(I)} (n)$.
Therefore the critical temperatures of composite string is given by

\begin{equation}
\beta_c^{(I)} = \pi \sqrt{\frac{D-2}{3Q(N)}}
= \frac{2}{N}\sqrt{\frac{2\pi}{T_{II}}}\,\,
\mbox{.}
\label{48}
\end{equation}

\begin{equation}
\beta_c^{(II)} = \pi \sqrt{\frac{D-2}{3Q(N)}}
= \frac{2}{N}\sqrt{\frac{2\pi}{T_{II}}}\,\,
\mbox{,}
\label{49}
\end{equation}
and $\beta_c^{(I)} = \beta_c^{(II)} = \beta_c$.

\vskip 1.0cm
\section{High Genera}

The aim of this section is to consider thermodynamic properties of the 
composite string to arbitrary genus ${\rm g}$ associated with Riemann surface 
world--sheet $\Sigma_{\rm g}$. Such considerations
allow us to identify the critical temperature at arbitrary loop order. 
It is  well-known that the genus-${\rm g}$ temperature contribution to the
free energy for the bosonic string can be written as \cite{clain87}

\begin{equation}
F_{\rm g}(\beta) = \sum_{{\bf m,n} \in {\bf Z}^{2{\rm g}}/\{{\bf 0}\}}
\int (d\tau)_{WP}\, (\det
P^+P)^{1/2}(\det
\Delta_{\rm g})^{-13}
e^{-\Delta S(\beta;{\bf m},{\bf n})}\,,
\label{50}
\end{equation}
where $(d\tau)_{WP}$ is the Weil--Petersson measure on the
Teichm\"{u}ller space. This measure as well as the factors $\det (P^+P)$ 
and $\det\Delta_{\rm g}$ are each individually modular invariant \cite{clain87}.
In addition,

\begin{equation}
I_{\rm g}(\tau)=(\det P^+P)^{1/2}(\det
\Delta_{\rm g})^{-13}=e^{c(2{\rm g}-2)}
\left(\frac{d}{ds}Z(s)|_{s=1}\right)^{-13}Z(2)\,,
\label{51}
\end{equation} 
where $Z(s)$ is the Selberg zeta function and $c$ an absolute
constant \cite{dhoker88}. Furthermore, the winding--number factor
has the form of a metric over the space of windings, namely

\begin{equation}
\Delta S(\beta;{\bf m},{\bf n})=
\frac{\pi}{2} NT_{II}\beta^2 [m_\ell\Omega_{\ell i}-n_i]
((\Im \Omega)^{-1})_{ij}
[\bar{\Omega}_{jk}m_k-n_j]={\rm g}^{\mu\nu}(\Omega){\cal N}_{\mu}{\cal N}_{\nu},
\label{52}
\end{equation}
\\
where $\mu,\nu = 1,2,..,2g$, 
$\{{\cal N}_1,...,{\cal N}_{2{\rm g}}\}
\equiv \{m_1,n_1,...,m_{\rm g},n_{\rm g}\}$.
The periodic matrix $\Omega$, corresponding to the string world--sheet of genus
${\rm g}$, is a holomorphic function of the moduli, $\Omega_{ij}=\Omega_{ji}$
and $\Im \Omega >0$. The matrix $\Omega$ admits a decomposition into
real symmetric ${\rm g} \otimes {\rm g}$ matrices: 
$\Omega =\Omega_1+i\Omega_2$. As a result

\begin{equation}
{\rm g}(\Omega_1+i\Omega_2)=\left(
\begin{array}{cc}
\Omega_1\Omega_2^{-1}\Omega_1+\Omega_2 & -\Omega_1\Omega_2^{-1}\\
-\Omega_2^{-1}\Omega_1 & \Omega_2^{-1}
\end{array}
\right).
\label{53}
\end{equation}
Besides, ${\rm g}(\Omega)=\hat{\Lambda}^t{\rm g}(\Lambda(\Omega))\hat{\Lambda}$ 
\cite{dhoker88}, where $\Lambda$ is an element of the symplectic modular group
$Sp(2{\rm g},{\bf Z})$ and the associated tranformation of the periodic matrix
reads $\Omega \mapsto \Omega'=\Lambda(\Omega)=(A\Omega+B)(C\Omega+D)^{-1}$. As a
consequence, the winding factor
$\sum _{{\bf m},{\bf n}}\exp{[-\Delta S(\beta;{\bf m},{\bf n})]}$
is also modular invariant.
It can be shown that the $2{\rm g}$ summations present in the expression
for $F_{\rm g}(\beta)$ can be replaced by a single summation together with a
change in the region of integration from the fundamental domain to the analogue
of the strip $S_{a_1}$ related to the cycle $a_1$, whose choice is
entirely arbitrary \cite{clain87,murphy89}. Then, one has

\begin{equation}
F_{\rm g}(\beta) = \sum^\infty_{r=1} \int (d\tau)_{WP}\, I_{\rm g}(\tau)
\exp \left\{-\frac{\pi}{2}NT_{II}\beta^2r^2 (\Omega_{1i}
((\Im \Omega)^{-1})_{ij} \bar{\Omega}_{j1})\right\}.
\label{54}
\end{equation}
To make use of the Mellin transform the genus--${\rm g}$ free energy can
be present in the form \cite{bytsenko93,elidrom,bytsenko96}

$$
F_g(\beta)=\frac{1}{2\pi i}\int_{\Re\,s=s_0}ds
\Gamma(s)\zeta(2s)\left(\frac{\pi}{2}NT_{II}\beta^2\right)^{-s}
$$

\begin{equation}
\times
\left\{\int(d\tau)_{WP}\, I_{\rm g}(\tau) [\Omega_{1i}
((\Im \Omega)^{-1})_{ij} \bar{\Omega}_{j1}]^{-s}\right\} _{({\rm Reg})}.
\label{55}
\end{equation}
In order to deal with Eq. (55) the integrals on a suitable variable in 
$(d\tau)_{WP}$ should be understood as the regularized ones. In this way the order 
of integration may be interchanged.

The critical behaviours of closed and open strings of the composite model coincide
(at least at level ${\rm g}=1$). 
Let us consider, for example, the open string genus-${\rm g}$ contribution to the 
free energy. The matrix $\Omega$ may be chosen as  
$\Omega={\rm diag}(\Omega_2,\Omega_2^{-1})$. In
the limit $\Omega_2 \rightarrow 0$, one has

\begin{equation}
\exp \left\{-\frac{\pi}{2}NT_{II}\beta^2 ({\bf {\cal N}}^t 
\Omega {\bf {\cal N}})^{-s}\right\} 
\longrightarrow 
\exp\left\{-\frac{\pi}{2}NT_{II} \beta^2 \Omega_2^{-1}{\bf {\cal N}}^t 
{\bf {\cal N}}\right\}\, ,
\label{56}
\end{equation}
and

\begin{equation}
\left(\sum_{{\bf {\cal N}} \in {\bf Z}^{2g}/\{\bf 0\}} ({\bf {\cal N}}^t \Omega 
{\bf {\cal N}})^{-s}\right)_{(\Omega_2\rightarrow 0)}
\longrightarrow \Omega_2^s
\sum_{{\bf {\cal N}} \in {\bf Z}^{\rm g}/\{\bf 0\}}
({\bf {\cal N}}^t {\bf {\cal N}}) ^{-s}=\Omega_2^s
Z_{\rm g}|^{\bf 0}_{\bf 0}|(2s) \, ,
\label{57}
\end{equation}
where the Epstein zeta function of order ${\rm g}$ is defined by

\begin{equation}
Z_{\rm g}|^{\bf b}_{\bf h}|(s)=\sum_{{\bf {\cal N}} \in {\bf Z}^{\rm g}/\{\bf 0\}}
[(n_1+b_1)^2+...+(n_{\rm g}+b_{\rm g})^2]^{-s/2} 
\exp {[2\pi i ({\bf {\cal N}}^t,\bf h)]}\, .
\label{58}
\end{equation}
The corresponding contribution is given by

\begin{equation}
\frac{1}{2\pi i}\int_{\Re\,s=s_0}ds\, 
\Gamma(s)\left(\frac{\pi}{2}NT_{II} \beta^2 \right)^{-s}
Z_{\rm g}|^{\bf 0}_{\bf 0}|(2s)
\left\{\int d\tau_{WP} \Omega_2^{s}I_{\rm g}(\tau)\right\} _{({\rm Reg})}.
\label{59}
\end{equation}

Since a tachyon is present in the spectrum, the total free energy
will be divergent, for any ${\rm g}$. The infrared divergence may be
regularized by means of a suitable cutoff parameter. This divergence
could be associated with pinching a cycle non homologous at zero. The 
behavior of the factor $(d\tau)_{WP}\, I_{\rm g}(\tau)$ is given by 
the Belavin--Knizhnik double--pole result and has a universal character
for any ${\rm g}$. It should also be noticed that this
divergence is $\beta -$ independent and the meromormphic structure is
similar to genus--one case. As a consequence, the whole genus dependence 
of the critical temperature is encoded in the Epstein zeta function
$ Z_{\rm g}|^{\bf 0}_{\bf 0}|(2s)$ (see for details Ref. \cite{bytsenko93}).

For this reason, we mention the asymptotic properties of function
$Z_{\rm g}|^{\bf 0}_{\bf 0}|(2s)$. The following result holds:

\medskip
\par \noindent
\begin{corollary} {\rm (Ref. \cite{bytsenko93})}

\begin{equation}
B_{\rm g}\equiv \lim _{\Re s\, \rightarrow +\infty}
\frac{Z_{\rm g}|^{\bf b}_{\bf 0}|(2s+2)}{Z_{\rm g}|^{\bf b}_{\bf 0}|(2s)}=
[(\hat{b}_1-\eta_1)^2+...+(\hat{b}_g-\eta_g)^2]^{-1 } ,
\label{60}
\end{equation}
where at least one of the $b_i$ is noninteger,
$\hat{b}_i=b_i-[b_i]$
with $[b_i] $ the noninteger (decimal) part of $b_i$ and

\begin{equation}
\eta_i=\left\{
\begin{array}{cc}
0 \, , &\hspace{1cm} 0\leq \hat{b}_i \leq 1/2, \\
1 \, , & \hspace{1cm} 1/2\leq \hat{b}_i < 1  .
\end{array} \right.
\label{61}
\end{equation}
Furthermore, if ${\bf b} = (0,0,...0)$, then $B_{\rm g}=1$.
\end{corollary}
\medskip
\medskip
As a consequence, the interactions of bosonic strings do not modify the 
critical temperature. However one can consider different linear real 
bundles over compact Riemann surfaces and spinorial structures on them. 
The procedure of evaluation of the free energy in terms of the path
integral over the metrics does not depend on whatever type of real
scalars are considered. This fact leads to new contributions to the 
genus--${\rm g}$ integral (50). 

One could  investigate the role of these 
contributions for the torus compactification \cite{dhoker88,lerche89,bayin96}. 
In this case, the sum in Eq. (50) should be taken over the vectors on the 
lattice on which some space dimensions are compactified. The half--lattice 
vectors can be labelled by the multiplets $(b_1,..,b_p)$, with $b_i=1/2$. 
The critical temperature related to the multiplet ${\bf b} = (b_1,...b_p,0,..,0)$ 
can be easily evaluated by means of Eq. (60), which gives $B_p=4p^{-1}$. 
As a result $\beta_{c,p}= (2/\sqrt p)\,\beta_c$.
As an example, we note the particular multiplets $(0,..,1/2,..,0)$ and 
$(0,..1/2, 1/2,..0)$, where only one $b_i$ and two $b_i$ are different from
zero. In this case we have ``minimal" critical temperatures given by 
$\beta_{c,1}^{-1}=\beta_c^{-1}/2$ and  $\beta_{c,2}^{-1}=\beta_c^{-1}/{\sqrt 2}$ 
respectively.

\section{Concluding Remarks}

Making use of the Meinardus theorem in Section 6, we found the critical 
temperatures of the two kinds of pieces in the string (I and II), to be equal, 
and to be given by Eqs.~(48) and (49) for arbitrary spacetime dimension $D$. 
The calculation generalizes earlier calculations, of the Casimir energy of the 
$2N-$ piece string in Ref.\cite{brevik97}, and of thermodynamics of the 2--piece 
string in Ref.\cite{brevik98}. Interactions of bosonic parts of a piecewice
uniform string do not modify the critical temperatures. However, for the sectors
of parts having a spinor structure, the critical temperatures, associated 
with genus ${\rm g}$, depend on the windings.

The following point ought finally to be stressed in order to prevent misunderstanding.
 The reason why we have dealt with the limit of extreme tensions, $x=T_I/T_{II} \rightarrow 0$, is merely
practical; therewith the eigenvalue branches degenerate into one single branch, Eq.~(13). This simplification is 
however not of fundamental importance; in particular, it has no bearing on the general problem of how to distinguish between the actions for massive and massless particles. We never put $T_I$ exactly equal to zero.  Thus in Eq.~(20) the denominator goes to zero when $T_I$ does, but the same $T_I$ is restored again in Eqs.~(22) and (26). The result is that the commutation rules for the creation and annihilation operators, Eq.~(29), take the same form as usual in quantum field theory.

\section{Acknowledgements}
The research of A.A. Bytsenko has been supported in part by the Russian
Foundation for Basic Research (grant No. 01--02--17157).

\end{document}